# Influence of experimental context on the development of anhedonia in male mice imposed to chronic social stress.

N.P. Bondar, I.L. Kovalenko, D.F. Avgustinovich, N.N. Kudryavtseva

Neurogenetics of Social Behavior Sector, Institute of Cytology and Genetics SD RAS, Novosibirsk, 630090, Russia, e-mail: natnik@bionet.nsc.ru

**Abstract**

Anhedonia is one of the key symptoms of depression in humans. Consumption of 1% sucrose solution supplemented with 0.2% vanillin was studied in two experimental contexts in male mice living under chronic social stress induced by daily experience of defeats in agonistic interactions and leading to development of depression. In the first experiment, vanillin sucrose solution was made available as an option of water during 10 days to mice living in group home cages. Then the mice were subjected to repeated social defeat stress and during stress exposure they were provided with both vanillin sucrose solution and water using a free two bottles choice paradigm. In the other experiment, vanillin sucrose solution were first offered to mice after 8 days of exposure to social defeat stress.

Males familiar with vanillin sucrose solution showed vanillin sucrose preference while experiencing defeat stress: consumption of vanillin sucrose solution was about 70% of total liquid consumption. However, the consumption of vanillin sucrose solution per gram of body weight in mice imposed to social stress during 20 days was significantly lower than in control males. In the second experiment, males after 8 days of social defeat stress were found to consume significantly less vanillin sucrose solution as compared with control males. On average during two weeks of measurements, vanillin sucrose solution intake was less than 20% of total liquid consumption in males. Consumption per gram of body weight also appeared to be significantly lower than in control group. Influence of the experimental context on the development of anhedonia, which was measured by the reduction in sucrose solution intake by chronically stressed male mice, has been discussed. It was assumed that the absence of attenuation of sucrose intake in stressed males is not essentially the sign that anhedonia and, consequently, depression are not being formed, at least in the context of animal models. At the same time, in animals a reduction in sucrose consumption could be indicative of depression only in combination with other symptoms characterizing this psychoemotional disorder.

## ВВЕДЕНИЕ

Одним из главных симптомов большой депрессии является ангедония, под которой понимается отказ от получения удовольствий людьми в результате развития болезни [APA, DSM-IV, 1994]. Для имитации депрессивно-подобного состояния у животных применяют непредсказуемый физический стресс, хронический слабый стресс, социальный стресс, различные виды комбинированного стресса, в результате чего у животных наблюдали снижение потребления сахарозы или сахарина в условиях свободного выбора наряду с водой [Katz, 1982; Monleon et al., 1995; Weiss, 1997; Willner, 1997; Moreau, 2002; Pothion et al., 2004; Strekalova et al., 2004; Rygula et al., 2005]. По аналогии с депрессивными людьми, полагают, что снижение потребления сахарозы животными свидетельствует о гедоническом дефиците (ангедонии), являющемся симптомом депрессии.

Ранее нами было показано, что под влиянием ежедневного стресса социальных поражений в агонистических взаимодействиях у самцов мышей развивается состояние депрессии, определяемой по множеству изменений в поведении, физиологическом состоянии и медиаторной активности мозга [Kudryavtseva et al., 1991; Kudryavtseva, Avgustinovich, 1998; Августинович и др., 2004], которые соответствовали таковым у депрессивных больных. Исследование возможного развития ангедонии у депрессивных самцов в ответ на предъявление предпочитаемого продукта, используемого в качестве гедонического стимула – сыра, выявило, однако, некоторые особенности пищевого поведения [Кудрявцева и др., 2006]. В упомянутом эксперименте мы вначале давали животным, проживавшим в домашних групповых клетках, сыр наравне с обычной едой (гранулами) в течение двух недель в условиях свободного выбора. После этого самцы подвергались воздействию социального стресса, в процессе которого им по-прежнему был предоставлен выбор между двумя продуктами. Было показано, что потребление сыра, вкус которого мышам был длительно знаком, даже несколько увеличивалось в количественном отношении у самцов мышей в процессе воздействия социального стресса. Это, казалось бы, свидетельствовало об отсутствии развития ангедонии. Однако после периода лишения сыра (депривации), депрессивные самцы продемонстрировали сниженное потребление по сравнению с контролем и сниженную пищевую мотивацию, оцениваемую по поведенческой реакции на сыр, помещенный на расстоянии и отделенный преградой. Снижение пищевой мотиации в отношении ранее предпочитаемого продукта были расценены нами как признак ангедонии у самцов мышей в результате развития депрессии. Было предположено, что феномен усиления потребления сыра в условиях стресса мог иметь место только в отношении сыра. В настоящем эксперименте было исследовано возможное развитие ангедонии в ответ на другой, традиционно использующийся для мышей и крыс гедонический стимул – раствор сахарозы в двух экспериментальных ситуациях. В первом случае, как и в эксперименте с сыром, вначале раствор сахарозы (с добавлением ванилина) был предоставлен групповым мышам в условиях свободного выбора наряду с водой в домашних клетках в течение 10 дней. Затем мыши были подвергнуты социальному стрессу, в процессе которого им по-прежнему был предоставлен выбор между раствором ванилиновой сахарозы и водой. В другой серии раствор ванилиновой сахарозы животным давали впервые после 8 дней социального стресса.

## МЕТОДИКА

Эксперименты проводили на половозрелых самцах мышей линии C57BL/6J в возрасте 3-4 мес. и массой тела 28-30 г. Животных разводили и содержали в стандартных условиях вивария Института цитологии и генетики СО РАН. Воду и корм (гранулы) самцы получали в достаточном количестве. Световой режим был равен 12:12 часам. До начала эксперимента мышей содержали по 6-8 особей в пластиковых клетках размером 36x23x14см. Все процедуры осуществляли в соответствии с международными правилами проведения экспериментов с животными (European Communities Council Directives of 24 November 1986, 86/609/EEC).

### *Формирование депрессии у самцов мышей*

Для формирования депрессии у самцов мышей использовали модель сенсорного контакта [Kudryavtseva et al., 1991]. Животных попарно помещали в экспериментальные клетки, разделенные пополам прозрачной перегородкой с отверстиями, позволявшей мышам видеть, слышать, воспринимать запахи друг друга (сенсорный контакт), но предотвращавшей физическое взаимодействие. Ежедневно во второй половине дня (15.00-17.00 часов) убирали перегородку, что проводило к межсамцовым конфронтациям. Во время первых 2-3 тестов выявляли победителей (агрессоров), и особей, терпящих поражения (побежденные, жертвы) при взаимодействии с одним и тем же партнером. В дальнейшем после теста побежденного самца пересаживали в новую клетку к незнакомому агрессивному партнеру, сидящему за перегородкой. Если интенсивные атаки со стороны нападающей особи во время агрессивных столкновений длились более 3-х минут, взаимодействие самцов прекращали, вновь устанавливая между ними перегородку. В других ситуациях тест продолжался 10 минут. Агрессоры в течение всего эксперимента проживали в своих отсеках. В качестве контроля использовали интактных самцов мышей, которых рассаживали в индивидуальные клетки на 5 дней, в течение которых снимается эффект групповых взаимодействий и не развивается эффект изоляции [Kudryavtseva, 1991]. Было экспериментально подтверждено, что такой контроль в условиях модели сенсорного контакта является наилучшим из всех возможных контролей [Августинович и др., 2005].

*Экспериментальные дизайны*

В нашем эксперименте был использован раствор 1% сахарозы с добавлением ванилина (0,2%), обладающего привлекательным запахом для мышей [Ducottet, Belzung, 2004]. В предварительных экспериментах было показано, что мыши, постоянно живущие в группах, пьют раствор ванилиновой сахарозы также хорошо, как и просто сахарозы. После замены раствора сахарозы на ванилиновую сахарозу и наоборот мыши предпочитали пить раствор ванилиновой сахарозы в большем количестве, чем просто раствор сахарозы. Это побудило нас остановить выбор на растворе ванилиновой сахарозы, обладающей более выраженными для мышей аттрактивными свойствами, возможно, благодаря приятному запаху и вкусу.

**Эксперимент 1.** Исследовали потребление раствора ванилиновой сахарозы самцами мышей с повторным опытом социальных поражений, имевших предварительный опыт знакомства с нею. Для этого мышам, проживавшим с момента отсадки от матерей в группах по 6-8 животных в стандартных клетках предлагали в течение 10-ти дней свободный выбор между двумя бутылочками с раствором ванилиновой сахарозы и водой. Ежедневно бутылочки взвешивали и меняли местами, чтобы избежать выработки предпочтения места у мышей. Потребление раствора ванилиновой сахарозы и воды групповыми животными оценивали в расчете на общую массу животных в клетке.

После 10-дневного периода знакомства с раствором ванилиновой сахарозы животных помещали в экспериментальные клетки, разделенные прозрачной перегородкой с отверстиями. Затем они участвовали в агонистических взаимодействиях в условиях модели сенсорного контакта в течение 20 дней. При этом на протяжении всего эксперимента самцам был предоставлен выбор между раствором ванилиновой сахарозы и водой. На 8-й день, когда альтернативные типы социального поведения (победители и побежденные, агрессоры и жертвы) были устойчиво сформированы, производили первый замер суточного потребления воды и раствора ванилиновой сахарозы, а также массы тела побежденных самцов. В последующем замеры проводили каждые 6 дней (на 8, 14, и 20 дни агонистических взаимодействий). Исследовали *потребление* раствора ванилиновой сахарозы, выраженное в граммах на массу тела животного, *потребление воды* и *общее потребление жидкости,* также выраженные в граммах на массу тела животного. Подсчитывали показатель *предпочтения* раствора ванилиновой сахарозы, выраженный в процентах от общего количества выпитой жидкости за сутки;

**Эксперимент 2.** Исследовали потребление раствора ванилиновой сахарозы самцами мышей с опытом социальных поражений, не знакомых прежде со вкусом раствора ванилиновой сахарозы. После 7 дней агонистических взаимодействий, побежденным самцам впервые давали раствор ванилиновой сахарозы на ночь (с 18 часов вечера до 10 часов утра, при этом бутылочку с водой убирали) для форсированного знакомства со вкусом раствора ванилиновой сахарозы. Утром животным дополнительно ставили воду. В последующем выбор между водой и раствором был предоставлен постоянно на протяжении всего эксперимента. Суточное потребление жидкостей исследовали каждые 6 дней. Как и в первом эксперименте, бутылочки ежедневно меняли местами, чтобы предотвратить у мышей привыкание к месту расположения жидкостей. Ежедневные агонистические взаимодействия продолжались в течение всего экспериментального периода. У контроля и у побежденных самцов, находившихся в условиях социального стресса, оценивали те же параметры потребления воды и раствора ванилиновой сахарозы, что и в первом эксперименте.

В обоих экспериментах сравнение показателей потребления жидкостей проводили по отношению к контрольным животным, находившимся в условиях индивидуального содержания 5 дней, в течение которых им был предоставлен выбор между водой и раствором ванилиновой сахарозы. Потребление обеих жидкостей контрольными самцами оценивали по средним значениям параметров за 4 дня, начиная со второго, предполагая, что в первый день происходит знакомство с раствором ванилиновой сахарозы, а в последующем животные уже демонстрируют предпочтение.

*Статистика.*

Статистическую обработку данных проводили с использованием ANOVA для повторных измерений количества потребляемого раствора ванилиновой сахарозы и воды. Сравнение показателей по группам проводили с помощью t-критерия Стьюдента для зависимых и независимых выборок.

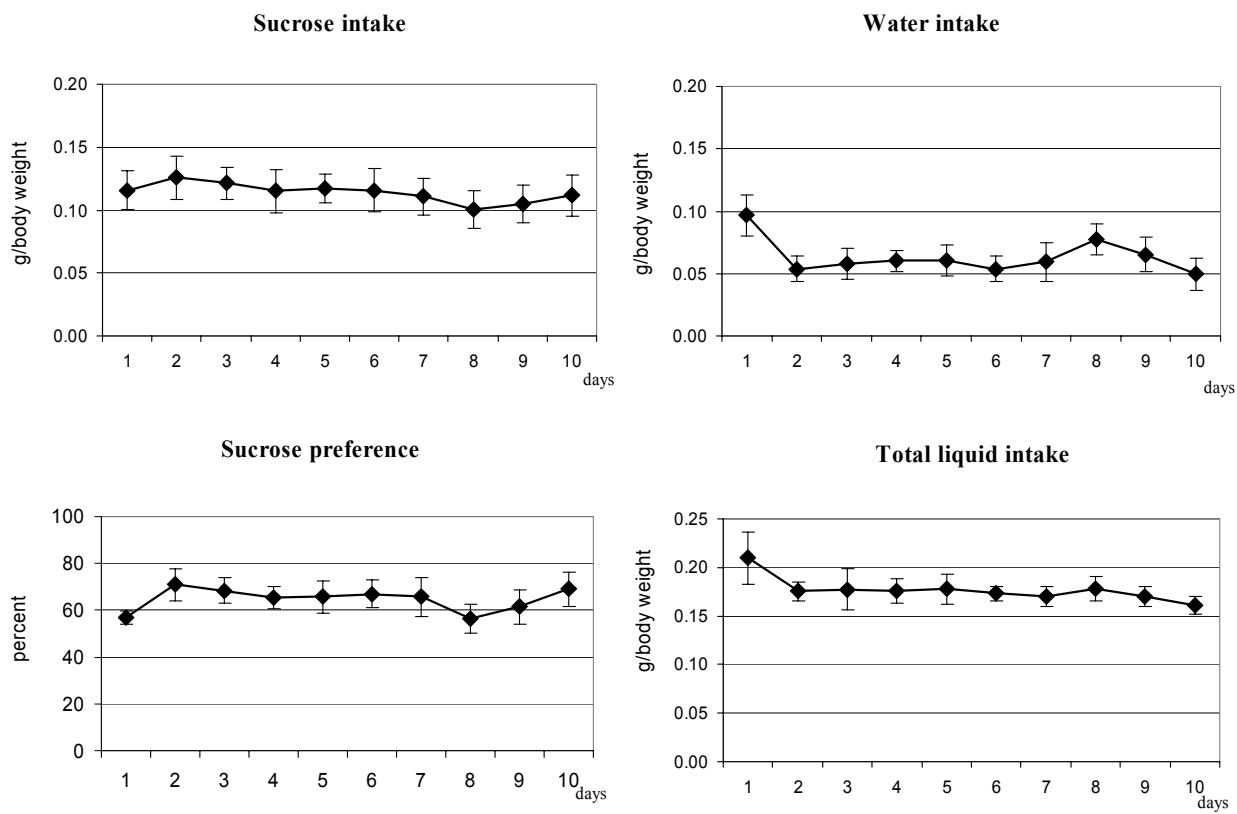

**Рисунок 1**. Потребление раствора ванилиновой сахарозы и воды в условиях свободного выбора групповыми животными. Voluntary vanillin sucrose solution and water intake by group-housed mice.

Различия между группами считали значимыми при p < 0.05. В группах было по 8-11 животных.

### РЕЗУЛЬТАТЫ

Самцы, проживавшие в групповых клетках, в первый день потребляли раствор ванилиновой сахарозы в количестве 57% от общего объема потребляемой жидкости. На второй день процент предпочтения увеличивался до 65-70% и сохранялся на этом уровне в течение 9 дней (Рис 1). Со второго по десятый день не было обнаружено влияния фактора повторных измерений на процент предпочтения раствора ванилиновой сахарозы [F(8,40) = 1.3; P>0.05], на ее потребление [F(8,40) = 1.5; P>0.05], на потребление воды [F(8,40) = 1,3; P>0.05] и общее потребление жидкости [F(8,40) = 0.8; P>0.05] на грамм массы тела (Рис. 1).

Сравнение средних значений потребления жидкостей между групповыми животными и контролем, не выявило различий по проценту предпочтения раствора ванилиновой сахарозы и по потреблению воды на грамм массы тела (P>0.05). Однако потребление раствора ванилиновой сахарозы и общее потребления жидкости на грамм массы тела у контрольных животных было существенно больше (P<0.05 и P<0.01, соответственно, Таблица). После помещения животных, длительно знакомых со вкусом раствора ванилиновой сахарозы, в условия социального стресса в **Эксперименте 1**, было отмечено сходное потребление раствора ванилиновой сахарозы самцами, проживавшими в группах, и самцами, впоследствии получившими опыт поражений - жертвами, по всем исследованным параметрам (P>0.05, данные не приведены). В то же время, по отношению к контролю (рис.2), у побежденных самцов, жертв, было достоверно ниже потребление раствора ванилиновой сахарозы (P<0.05 для точки 1 и 2, P<0.01 для точки 3), и общее потребление жидкости (P<0.01 для точки 1 и 2, P<0.001 для точки 3). Количество выпитой воды на грамм веса

**Таблица.** Потребление раствора ванилиновой сахарозы и воды самцами мышей, содержавшимися в группе и индивидуально (контроль)
Voluntary vanillin sucrose solution and water intake by group- and single- (control) housed male mice

| Parameters | Group mice | Control |
|---|---|---|
| **Sucrose preference, %** | 65.5±5.2 | 74.5±6.1 |
| **Sucrose intake, g/body weight** | 0.114±0.013 | 0.157±0.013* |
| **ater intake, g/body weight** | 0.060±0.010 | 0.051±0.012 |
| **Total liquid intake, g/body weight** | 0.173±0.011 | 0.209±0.007** |

\* - p<0.05, \*\* - p<0.01 **vs** group-housed mice.

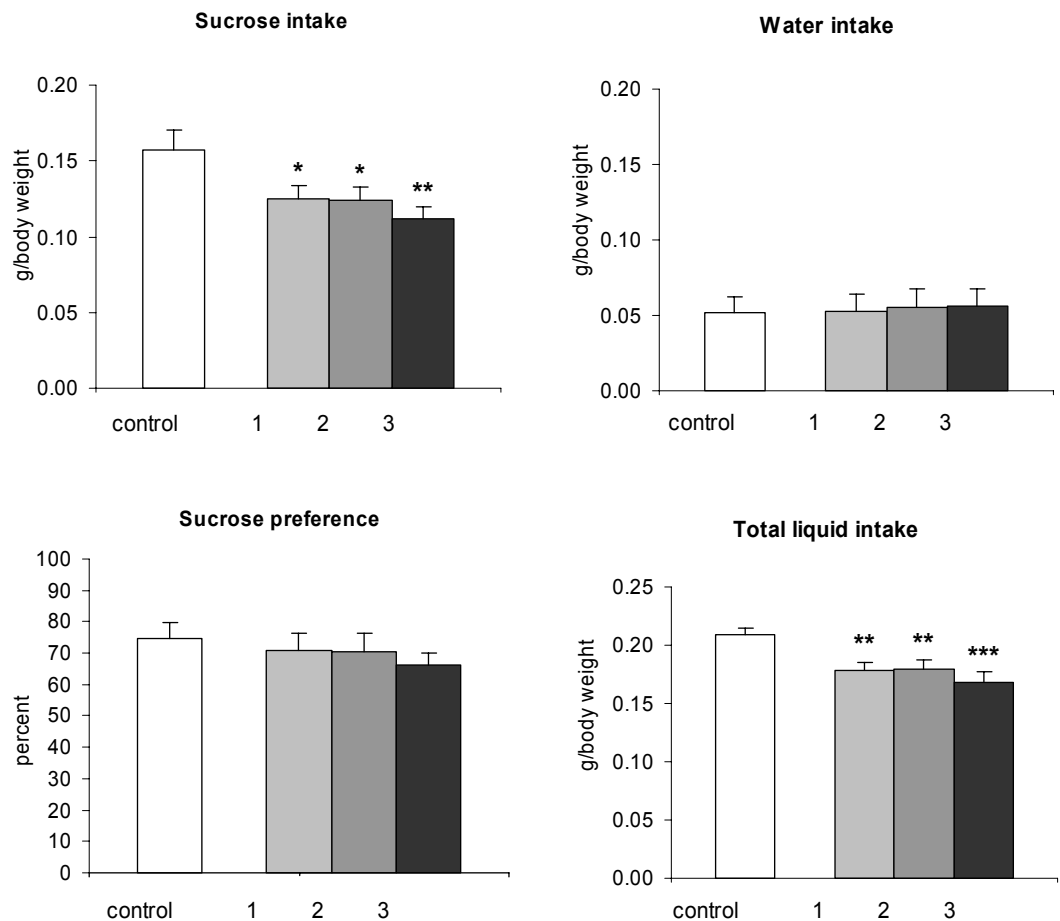

**Рисунок 2**. Потребление раствора ванилиновой сахарозы и воды самцами мышей, имевшими предварительный опыт знакомства с раствором ванилиновой сахарозы в течение 10 дней и затем помещенных в условия социального стресса 1, 2, 3 – точки измерения на 8, 14, 20 дни агонистических взаимодействий, соответственно. * - P< 0.05, ** - P<0.01, *- P<0.001 - по сравнению с контролем.
Voluntary vanillin sucrose solution and water intake by male mice familiar with vanillin sucrose solution during 10 days and then placed in the condition of daily chronic social stress. 1, 2, 3 - measures on 8, 14, 20 day of agonistic interactions, respectively. * - P< 0.05, ** - P<0.01, *- P<0.001 - vs control.

животного и процент предпочтения раствора внилиновой сахарозы у этих групп животных во всех точках измерения достоверно не отличались (P>0.05) (Рис.2).

В **Эксперименте 2** при первом измерении ванилиновой сахарозы на 8-й день конфронтаций жертвы выпивали 26,5% раствора ванилиновой сахарозы от общего количества выпитой жидкости (Рис.3). По сравнению с контролем у них были существенно ниже показатели процента предпочтения раствора ванилиновой сахарозы (P<0.001, для всех точек), потребления раствора ванилиновой сахарозы (P<0.001, для всех точек) и воды (P<0.001, для всех точек). Было обнаружено влияние фактора повторных измерений на показатель общего потребления жидкости [F(2,14) = 4.6, P<0.05]. Достоверное снижение общего количества потребленной жидкости было найдено в 3-й точке измерения по сравнению с 1-й (P<0.05) и 2-й (P<0.01) точками измерения. Кроме этого, в 3-й точке было найдено снижение потребления воды (P<0.01) по сравнению со 2-й точкой.

**ОБСУЖДЕНИЕ**

Исследования показали, что у мышей может быть сформировано гедоническое поведение по отношению к раствору ванилиновой сахарозы. Очевидно, что раствор ванилина и сахарозы более предпочитаем мышами по вкусовым качествам, чем вода. Возможно, быстрое приучение обусловлено запахом ванили, являющимся для мышей аттрактивным. Уже на второй день предъявления раствора ванилиновой сахарозы самцы, проживавшие в группах и индивидуально выходили на максимальный уровень потребления раствора – около 70% от общего количества выпитой жидкости. Групповые животные устойчиво предпочитали раствор ванилиновой сахарозы воде в течение 9 дней. Однако потребление раствора ванилиновой сахарозы на грамм суммарной массы животных в группах было существенно ниже по сравнению с контрольными самцами, находив-

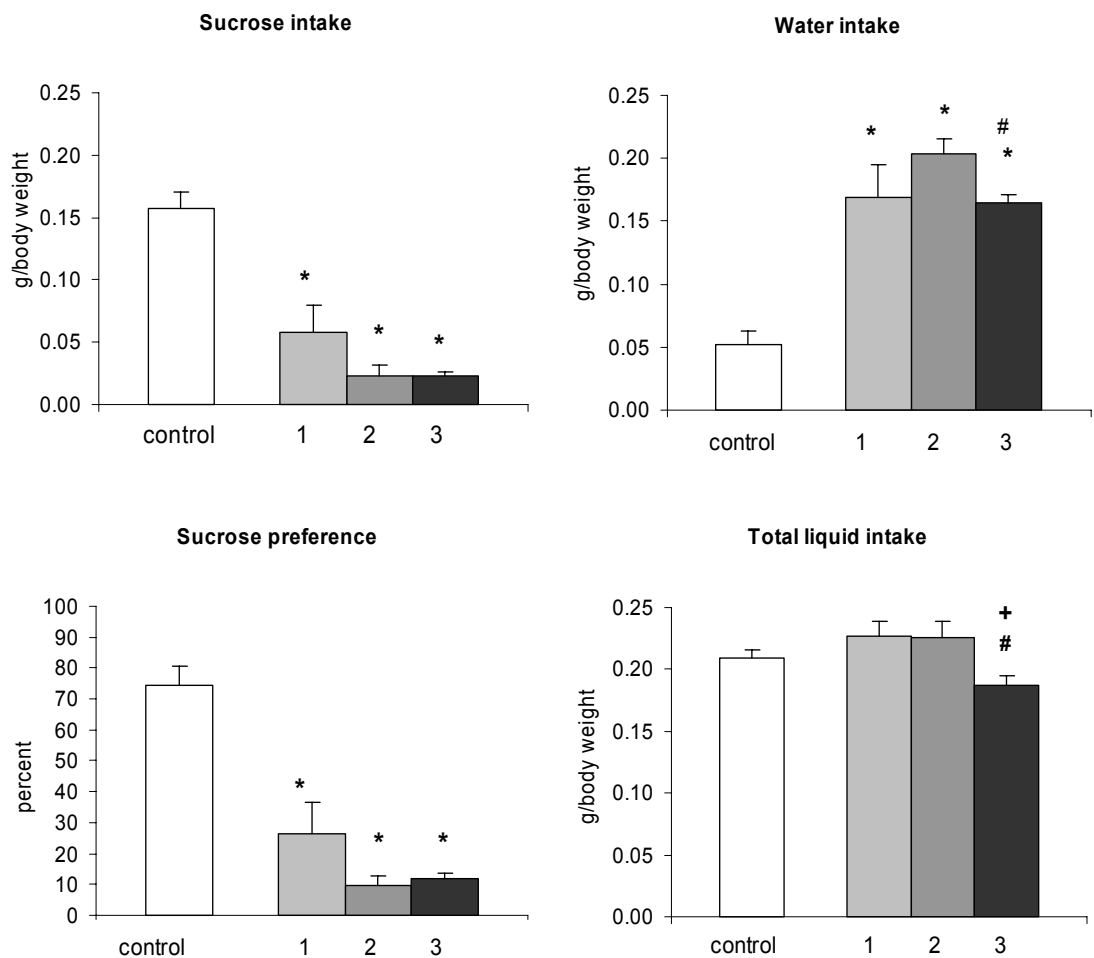

**Рисунок 3**. Потребление раствора ванилиновой сахарозы и воды самцами мышей, находившимся в условиях социального стресса и не имевших предварительный опыт знакомства с раствором ванилиновой сахарозы. 1, 2, 3 – точки измерения на 8, 14, 20 дни агонистических взаимодействий, соответственно. * - P<0.001 - по сравнению с контролем, # - P<0.01 - по сравнению со 2-й точкой измерения, + - P<0.05 - по сравнению с 1 точкой измерения.

Voluntary vanillin sucrose solution and water intake by male mice placed in the conditions of daily chronic social stress and non-familiar with vanillin sucrose solution. 1, 2, 3 - measures on 8, 14, 20 day of agonistic interactions, respectively.* - P<0.001 - vs control, # - P<0.01 - vs 2nd measure, + - P<0.05 – vs 1st measure.

шимся в индивидуальных клетках. Как следствие, у групповых мышей было снижено и общее потребление жидкости. Можно полагать, что сниженное потребление раствора ванилиновой сахарозы у самцов мышей является следствием иерархических отношений в группах, в результате которых не все особи в равной степени имеют доступ к раствору, являющемуся источником удовольствия. Кроме того, можно думать, что подчиненные самцы, каковыми в популяциях лабораторных мышей являются все кроме одного – доминанта [Mackintosh, 1970], проявляют меньший интерес к новому стимулу – раствору ванилиновой сахарозы. В этом случае, сниженное потребление раствора ванилиновой сахарозы по сравнению с контролем может являться признаком ангедонии, развивающейся у большинства подчиненных самцов в условиях клеточного содержания в группах.

Было выявлено существенное влияние экспериментального контекста на потребление раствора ванилиновой сахарозы самцами мышей, находившимся в условиях социального стресса. Самцы, длительно знакомые со вкусом предпочитаемой жидкости, предпочитали ее пить и на фоне стресса социальных поражений: процент потребления раствора ванилиновой сахарозы у самцов в первые и последующие дни конфронтаций составлял около 70% от общего количества выпитой жидкости. Однако потребление раствора ванилиновой сахарозы в расчете на массу тела животного было достоверно ниже по сравнению с контролем, при сходном потреблении воды. Можно предположить, что уровень потребления раствора ванилиновой сахарозы побежденными самцами может отражать сформированную привычку потреблять сахарозу именно в том количестве, в котором они потребляли ее ранее при проживании в группах. На фоне социального стресса уровень привычного потребления не изменяется. Сходные данные были нами получены в отношения другого предпочита-емого продукта – сыра. Было показано, что в условиях свободного выбора в сочетании со стандартным кормом (гранулами) самцы с повторным опытом социальных поражений предпочитали потреблять сыр - 80% от общего количества еды - и на фоне социальных конфронтаций [Кудрявцева и др., 2006].

Во втором эксперименте первое предъявление раствора ванилиновой сахарозы производили на 8-й день социальных взаимодействий, когда побежденный статус самца был устойчиво сформирован под влиянием опыта социальных поражений. Оказалось, что уже в первые сутки и в последующем жертвы выпивали раствора ванилиновой сахарозы существенно меньше, чем контрольные особи и не предпочитали его воде – менее 20% от общего количества выпитой жидкости.

Это сниженное потребление раствора ванилиновой сахарозы, хорошо укладывается в картину развития депрессии, выявленную нами ранее у мышей, помещенных в условия хронического социального стресса [Kudryavtseva et al., 1991]. Развивающееся депрессивно-подобное состояние у побежденных мышей удовлетворяло всем критериям сходства с аналогичной психопатологией у людей: сходство этиологии, симптоматики, чувствительности к антидепрессантам, сходство нейрохимических изменений в мозге. У животных после 20-30 дней социального стресса, вызванного опытом социальных поражений, развивается поведенческий дефицит, сказывающийся на индивидуальном и социальном поведении в различных ситуациях, выявляется тревожность, наблюдаются изменения в соматическом состоянии мышей (снижение веса, уровня тестостерона и др.) [Kudryavtseva et al., 1991; Kudryavtseva, Avgustinovich, 1998]. Установлено вовлечение серотонергической системы мозга в механизмы развития депрессии [обзор, Августинович и др., 2004]. Традиционные анксиолитики и антидепрессанты оказывали положительные эффекты на состояние депрессивных самцов мышей [Августинович и др., 2004]. Однако оставался не исследованным один из главных симптомов, сопровождающий развитие депрессии у людей – это нарушение гедонического поведения в отношении витальных потребностей, имеющих место у здоровых животных. Первые эксперименты с сыром давали основание предполагать, что у депрессивных самцов мышей развивается ангедония [Кудрявцева и др., 2006]. В данной работе продемонстрировано снижение потребления предпочитаемого продукта и в отношении раствора ванилиновой сахарозы. Особенно выраженно это наблюдалось у стрессированных самцов, не знакомых ранее со вкусом сахарозы. Поскольку сниженное потребление раствора сахарозы у особей в стрессирующих условиях [Monleon et al., 1995; Strekalova et al. 2004; Pothion et al. 2004] рассматривают как проявление ангедонии, то животные, демонстрирующие сниженное потребление раствора сахарозы, считаются депрессивными. По сути получено дополнительное подтверждение развития депрессии, вызываемой повторным опытом социальных поражений у мышей в условиях модели сенсорного контакта.

Почему же потребление раствора ванилиновой сахарозы в условиях социального стресса значительно различается у самцов мышей в двух экспериментальных контекстах – с предварительным знакомством со вкусом и запахом раствора ванилиновой сахарозы и без оного? Можно предположить, что в первом случае раствор ванилиновой сахарозы становится обычным элементом пищи, которая нравится мышам в большей степени, чем вода. Даже в условиях социального стресса этот раствор остается предпочитаемым и не надоедает мышам.

У животных (не знакомых со вкусом сахарозы) после недельного социального стресса сниженное потребление раствора ванильной сахарозы наблюдалось уже при первом предъявлении. В процессе развития депрессии самцы еще больше снижали потребление раствора. Можно сказать, что они не хотят пить раствор ванилиновой сахарозы (в отличие от контроля), предпочитая пить воду. К третьей точке измерения они снижают и потребление воды, что свидетельствует о глубокой степени депрессивного состояния у таких самцов. Возникает вопрос, какие механизмы лежат в основе отказа депрессивных самцов от потребления раствора ванилиновой сахарозы в условиях социального стресса?

Полагают, что вкусовая чувствительность (ранее считавшаяся постоянным параметром генетического происхождения) может изменяться при сменах настроения [Heath et al., 2006]. Показано, что при депрессии у больных могут нарушаться процессы обоняния и вкусовые ощущения [Pause et al., 2001; Heckmann, Lang, 2006; Lombion-Pouthier et al., 2006], восстанавливающиеся после лечения антидепрессантами [Pause et al., 2001; Heath et al., 2006]. Поэтому факт очень малого потребления раствора ванилиновой сахарозы депрессивными самцами мог бы свидетельствовать о развитии аверсии (или о снижении чувствительности) по отношению к запаху или вкусу сахарозы, сходного с тем, что наблюдается у депрессивных людей. Можно также говорить как об отказе от удовольствия (ангедонии), так и об отсутствии интереса к новым стимулам и развитии индифферентности, отмечаемой нами ранее и в других ситуациях у депрессивных самцов мышей [Kudryavtseva, Avgustinovich, 1998].

В дополнение к сказанному хотелось бы рассмотреть несколько шире вопрос о снижении реакции на предпочитаемый пищевой продукт (сахароза или сыр), как показатель ангедонии у животных. Существуют достаточно много литературных данных, которыми показано, что нарушения в восприятии запаха и вкуса у людей наблюдаются при самых различных заболеваниях, например, при болезни Альцгеймера [Eibenstein et al., 2005], множественном склерозе [Zivadinov et al., 1999] и при многих других неврологических заболеваниях [Heckmann, Lang, 2006]. В то же время, у больных с выраженной депрессией может и не быть нарушений в каких-либо показателях вкусовых или обонятельных ощущений по сравнению со здоровыми людьми [Steiner et al., 1993; Thomas et al., 2002; Scinska et al., 2004]. Это значит, что выраженная депрессивная симптоматика у людей не всегда сопровождается изменениями в пищевом поведении. В то же время обнаруженные изменения могут и не являться специфическими именно для депрессии. Более того, под влиянием некоторых видов стресса иногда наблюдается повышенное потребление сладкого продукта у животных и человека [Willner et al., 1998; Murison, Hansen, 2001; Pijlman et al., 2003]. Утверждение, что ангедония является основным симптомом депрессии, обычно понимается в более широком смысле у людей – как отсутствие интереса к положительному подкреплению в разных ситуациях. Это может проявляться очень индивидуально и зависеть от жизненной истории индивида, его привычек, характера, нейрофизиологических особенностей, стадии депрессии и тд. Как показали наши эксперименты, снижение интереса к гедоническому стимулу (например, к раствору сахарозы), может проявляться у депрессивных животных не всегда или не очень выраженно, сильно зависит от контекста ситуации, в которой происходит знакомство с предпочитаемым продуктом. И потому, отсутствие снижения потребления раствора сахарозы животными, подвергшимся стрессирующему воздействию, показанное в нескольких работах [Weiss,1997; Harris et al., 1998; Nielsen et al., 2000], может и не являться определяющим признаком отсутствия депрессии у животных. В то же время снижение потребления раствора сахарозы стрессированными животными только вкупе с другими симптомами, характерными для этого психоэмоционального расстройства, может свидетельствовать о развитии депрессии. Как полагают [McKinney, Bunney, 1969], экспериментальная модель депрессии не может диагностироваться по ограниченному числу признаков и должна удовлетворять многим критериям сходства с этим заболеванием у людей.

### Заключение

Было показано, что контекст экспериментальной ситуации, в которой происходит знакомство с раствором ванилиновой сахарозы (предварительное знакомство или его отсутствие), имеет значение для развития нарушений гедонического поведения и выраженности ангедонии у самцов мышей, находящихся в условиях хронического социального стресса, который сопровождается развитием депрессии. У депрессивных самцов происходит развитие ангедонии, которая может проявляться как в том, что они снижают потребление раствора ванилиновой сахарозы, так и в том, что не проявляют интереса к ней вследствие развития индифферентности.